\documentclass[%
 reprint, twocolumn, %linenumbers,
 amsmath,amssymb,
 aps, physrev,
floatfix
]{revtex4-2}

\usepackage{graphicx}% Include figure files
\usepackage{dcolumn}% Align table columns on decimal point
\usepackage{bm}% bold math
\usepackage[colorlinks=true]{hyperref}% add hypertext capabilities
\usepackage{xcolor}
\usepackage{physics}

\begin{document}

\title{Quantum control of the environment in open quantum systems \\ enables rapid qubit reset}

\author{Carlos Ortega-Taberner}
\author{Eoin O'Neill}
\author{Paul Eastham}
\affiliation{School of Physics, Trinity College Dublin, Dublin 2, Ireland}
\affiliation{Trinity Quantum Alliance, Unit 16, Trinity Technology and Enterprise Centre, Pearse Street, Dublin 2, Ireland}

\date{\today}
\begin{abstract}
Qubit reset is crucial in quantum technology and is typically achieved by coupling the qubit to a dissipative environment. However, the achievable speed and fidelity are limited by qubit-environment entanglement. We use exact tensor-network simulations and a time-dependent variational approach to investigate these effects for transmon qubits with a time-dependent system-environment coupling. We show that they are due to the formation of a polaron state and how this can be reversed using a time-dependent coupling. Coupling protocols are identified which achieve reset with an excited-state population of $10^{-6}$ in $10$ ns. A related paper [C. Ortega-Taberner, E. O'Neill and P. R. Eastham, arXiv:XXXX.XXXX] addresses the complementary case of control via a time-dependent Hamiltonian. Our work shows how the dynamics of the environment of an open quantum system can be controlled to design effective quantum processes in non-Markovian systems.
\end{abstract}

\maketitle

The control of open quantum systems using applied fields underpins quantum science and technology~\cite{koch2016,koch2022}, enabling state preparation and stabilization \cite{pflanzer2013,ticozzi2014,albert2016,reiter2016,grimm2020,ma2021}, mitigation of decoherence \cite{bylander2011,lidar2014,wang2014,roszak2015,paulisch2016,lei2017,pokharel2018,ezzell2023} and control of thermodynamic processes \cite{mukherjee2013,ticozzi2014,mukherjee2015,dann2020,xu2022,erdman2022,khait2022,erdman2023,shubrook_numerically_2025}. It has been extensively studied assuming weak-coupling to a memoryless or Markovian environment. Going beyond such approximations is necessary to achieve accurate modelling and expose new regimes~\cite{devega2017}, but difficult because it amounts to solving a time-dependent many-body problem. Nonetheless, in the last few years powerful techniques for computing the dynamics of open quantum systems have been developed~\cite{chin2025,ortega-taberner2024} based on tensor networks~\cite{fux2024,link2024,strathearn2018}, which have allowed the discovery of optimal control protocols~\cite{butler2024,fux2021}.

Quantum control is usually considered in terms of a reduced description obtained by tracing out the environment, giving a restricted perspective in which the applied fields control the system. Here we show that once one goes beyond the Markovian approximation it becomes possible to control not just the system, but also the environment and the system-environment entanglement. Such entanglement is the origin of decoherence and classicality, giving our results fundamental as well as practical significance. We consider, as an important example, dissipative reset in superconducting qubits~\cite{basilewitsch2019,vaaranta2022,diniz2023,yuan2023,gautier2024,liu2024a}, where it has been predicted that there are intrinsic limits on speed and fidelity arising from non-Markovian effects~\cite{tuorila2019}. We use tensor network simulations to show that these limits arise from entanglement between the qubit and environment and are accurately captured by a polaron ansatz~\cite{silbey1984,xu2016}. Generalizing this ansatz using the time-dependent variational principle~\cite{Yao2013, Bera2016, Wang2016,Chen2018,Zhao2022,hackl2020} allows us to investigate the effect of time-dependent system-environment coupling and determine optimal switching protocols. These protocols overcome the predicted limitations and achieve fast high-fidelity reset, allowing reset to the environment temperature with residual populations below $10^{-5}$ in $10\; \mathrm{ns}$. Such fast high-fidelity reset could enable rapid cycle times, reusable ancillas~\cite{puente2024}, fast calibration~\cite{berritta2025} and error correction protocols~\cite{schindler2011,fowler2012, geher2025}. A related paper~\cite{longpaper} treats complementary aspects of this problem, including control of system-environment entanglement via the qubit Hamiltonian, reservoir engineering, and the impact of higher excited states.

\textit{Formation of the polaron during reset}.--- 
In a dissipative reset process a qubit is coupled to a dissipative environment, such as a resistor or transmission line, that relaxes it towards the ground state. Fast reset can be achieved by increasing the coupling to this environment, but this would be detrimental to normal operation. To avoid this, a tunable qubit-environment coupling can be used to switch the coupling off during normal operation and on during reset. An example is the quantum-circuit refrigerator~\cite{tan2017,sevriuk2019, sevriuk2022, Yoshioka2023, Hsu2020}, where the coupling can be switched in less than a nanosecond. 

To describe reset we use the spin-boson model~\cite{nitzan2006a,leggett1987},
\begin{align}
    H = \frac{\omega_q}{2}\sigma_x + \sum_k \omega_k b_k^\dagger b_k + u(t)\frac{\sigma_z}{2}\sum_k g_k (b_k^\dagger + b_k), \label{eq:sbm}
\end{align} with the qubit, of angular frequency $\omega_q$, treated as a two-level system. It is coupled to an environment of harmonic oscillators characterized by the spectral density $J(\omega) = \sum_k \abs{g_k}^2 \delta(\omega-\omega_k)$, which we take to be Ohmic with an exponential cutoff, $J(\omega)= 2\alpha \omega e^{-\omega/\omega_c}$. We take a typical transmon frequency $\omega_q/2\pi=5\;\mathrm{GHz}$, cutoff $\omega_c=\omega_q$, and coupling strength $\alpha=0.03$. The factor $u(t)$ corresponds to an overall control over the system-environment coupling strength. An active reset process begins with the qubit in the maximally-mixed state, the total system in a product state, $\rho=(I/2)\otimes\rho_{env}$, and no coupling $u=0$. The coupling is then switched on, $u=1$, to allow relaxation, and subsequently off, to avoid dissipation during normal operation. 

\begin{figure}[t]
\includegraphics[width=\columnwidth]{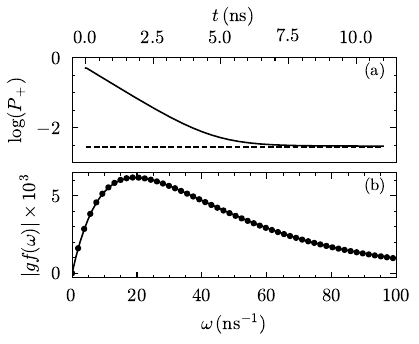}
\centering
\caption{ (a) Excited state population $P_+$ for reset in the spin-boson model (solid), showing exponential relaxation at early times and a non-zero residual population. This population corresponds to that predicted by the polaron grounnd state (dashed). (b) Spectral distribution of the bath displacements at $t=10\;\mathrm{ns}$ for the spin-boson model (solid) and corresponding displacements from the polaron ground state (circles).}
\label{fig:fig1}
\end{figure}

We begin by considering the case of instantaneous switches, and suppose the environment is at zero temperature. Within the Born-Markov approximation the
excited-state population decays exponentially to zero, allowing arbitrarily good reset. However, going beyond this approximation reveals that the population at late times is non-zero due to the build up of qubit-environment entanglement~\cite{tuorila2019}. This can be seen in Fig.~\ref{fig:fig1}(a), which shows the exact dynamics of the excited state population computed using the process-tensor version of the time-evolving matrix product operator algorithm (TEMPO) implemented in OQuPy~\cite{fux2024}. It leads to a trade-off between fidelity and speed, with weaker system-environment coupling giving a smaller residual population but a slower relaxation. 

This behavior can be explained by assuming that at late times the state approaches the ground-state of the coupled system, rather than the uncoupled one. In the weak coupling regime relevant here the former is accurately described by the polaron ansatz \cite{silbey1984,xu2016}
\begin{align}
    \ket{\Psi(\boldsymbol{{f}})} = \frac{1}{\sqrt{2}} \left(\ket{\uparrow,\boldsymbol{f}} - \ket{\downarrow,-\boldsymbol{f}} \right), \label{eq:polaron}
\end{align} where $\ket{\boldsymbol{f}}$ are coherent states of the bath oscillators with phase-space locations $\boldsymbol{f}$, formed by displacement of their ground states, $\ket{ \boldsymbol{f}} = \exp[ \sum_k (f_k b_k^\dagger - f_k^* b_k)]\ket{0}.$ Using a variational approach one finds the displacements obey \begin{align}f_k = -\frac{g_k}{2(\omega_q e^{-2\sum_k \abs{f_k}^2}+\omega_k)}\approx -\frac{g_k}{2(\omega_q +\omega_k)}\label{eq:polarondisps}\end{align} which appear in the qubit excited-state population \begin{equation} P_+=\left(1-e^{-2 \sum_k \abs{f_k}^2}\right)/2.\label{eq:pplus}\end{equation} This population is shown with the dashed line in Fig.~\ref{fig:fig1}(a), and can be seen to be in excellent agreement with the numerical result. 

To further probe the nature of the long-time limit we note that the polaron state ~\eqref{eq:polaron} is characterized by qubit-environment correlations such that there is a non-zero expectation value $g_k\langle \sigma_z(t) b_k(t)\rangle=g_k f_k$. This expectation value can be computed from the process tensor, using the approach of~\cite{gribben2022a} to relate it to the two-time correlation function $\langle \sigma_z(t)\sigma_z(t^\prime)\rangle$. 
As shown in Fig.~\ref{fig:fig1}(b), the qubit-environment correlations computed in this way, in the relaxed state at $t=10\,\mathrm{ns}$, are in perfect agreement with those in the polaron ground state.

\textit{Reversing polaron formation with decoupling}.--- 
To reduce the residual population in the qubit we now consider the effect of switching off the coupling $u(t)$ over a non-zero time interval. To treat this we exploit the accuracy of the polaron ansatz in conjunction with the time-dependent variational principle (TDVP) \cite{hackl2020}. We promote the bath displacements to time-dependent variables and use the TDVP (see end matter) to obtain the equations of motion 
\begin{align}\label{eq:eom}
    \dot{f}_k =& i f_k (\omega_q e^{-2\sum_k \abs{f_k}^2} + \omega_k)  + i\frac{1}{2}g_k u(t) \nonumber \\
    \approx&  i f_k (\omega_q + \omega_k)  + i\frac{1}{2}g_k u(t),
\end{align}
where in the weak coupling limit we can neglect the exponential, which decouples all equations of motion. 

\begin{figure}[t]
\includegraphics[width=\columnwidth]{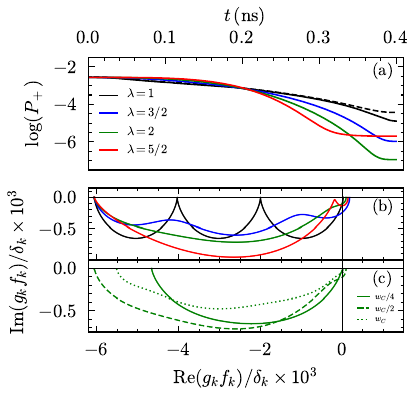}
\centering
\caption{ (a) Evolution of the residual population of the qubit during decoupling, obtained using TDVP for different values of the smoothness constant $\lambda$. For the linear case, $\lambda = 1$, the result of a TEMPO calculation is shown (dashed black). (b) Trajectory of the bath displacements during decoupling, for $\omega_k = \omega_C/2$ and different decoupling functions. (c) Displacements for $\lambda = 2$ and different bath frequencies. }
\label{fig:fig2}
\end{figure}

To validate \eqref{eq:eom} we compare the results to the exact dynamics of the time-dependent spin-boson model~\eqref{eq:sbm}. These are obtained using a modification of the TEMPO algorithm that allows for time-dependent couplings (see end matter). We first consider the dynamics starting from the relaxed polaron state at $u(t=0)=1$, and take a linear switch of the coupling to zero, $u(t)=1-t/t_f$, over a time interval $t_f=0.4\; \mathrm{ns}$, consistent with recent experiments using a quantum circuit refrigerator \cite{sevriuk2019}. Fig.~\ref{fig:fig2}(a) shows the excited-state populations from TEMPO (solid black) and the TDVP (dashed black). We observe that switching off the interaction to the environment in a finite time results in an improvement of the reset, due to weakening of the polaron, by two orders of magnitude. Furthermore, we observe a very good agreement in the dynamics obtained from both methods, which motivates the use of TDVP in the remainder of this work.

Decoupling can be further improved by analyzing the dynamics of the bath. Integrating the TDVP equations of motion we obtain
\begin{align}
    f_k(t) = f_{k0}e^{i\omega_k' t} + i\frac{1}{2}g_k e^{i\omega_k' t}\int_0^t dt' \,  e^{-i\omega_k' t'}u(t'),
\end{align}
where we introduced the renormalized bath frequencies $\omega_k' = \omega_k + \omega_q$ and the equilibrium point of the oscillators $f_{k0} = -g_k/2\omega_k'$. Integrating by parts, and noting that at the end of the process we have $u(t_f)=0$, gives the displacements 
\begin{align}
    f_k(t_f) = f_{k0}\int_0^{t_f} dt' \, e^{-i\omega_k' (t'-t_f)} \dot{u}(t').
\end{align}
Perfect reset occurs if $f_k(t_f) = 0$, and the integral above corresponds to the non-adiabatic contribution to the residual population at the end of the decoupling. Something to note from this expression is that the relevant control to study is actually the qubit-bath coupling, proportional to $u(t)$, and not the coupling strength, $\alpha$, which is quadratic on the couplings.

Repeated integration by parts gives an asymptotic expansion for $\omega_k^\prime t_f \gg 1$,
\begin{align}
    f_k(t_f)= f_{k0}\sum_{j=1}^\infty\frac{1-(-1)^je^{i\omega_k' t_f}}{(i\omega_k')^j} u^{(j)}(0),
\end{align} where we assume a symmetric switching function $u(t) = 1-u(t_f-t)$. We see that the final displacements can be reduced, and hence the fidelity improved, by suppressing the low-order derivatives of the switching function at the beginning and end of the protocol. However, this comes at the cost of increasing higher derivatives, which eventually outscale the frequency in the denominator and lead to a worse result. As an example, we consider the switching function
\begin{align}
    u(t) = 1-\frac{t^\lambda}{t^\lambda+(t_f-t)^\lambda},
\end{align} for which $u^{(j)}(0)=0$ when $j<\lambda$. Fig.~\ref{fig:fig2}(a) shows the excited state populations, computed for several values of $\lambda$. As discussed above linear decoupling, which corresponds to $\lambda = 1$, reduces residual population by two orders of magnitude. This is improved by $\lambda = 3/2$ and $\lambda = 2$ reaching over four orders of magnitude reduction in the excited population, $P_+ \approx 10^{-6.5}$. Above this, $\lambda = 5/2$ performs worse, due to the increase of higher derivatives. The origins of these effects can be seen in the trajectories of the displacements shown in Fig.~\ref{fig:fig2}(b,c). Fig.~\ref{fig:fig2}(b) shows the displacements of the oscillators with frequency $\omega_k=\omega_c/2$ for the different $\lambda$;  the $\lambda=2$ trajectory shows the smoothest behavior, while the others show oscillations induced by the switching. Importantly, the smooth switches lead to small final displacements across all bath frequencies, as shown in Fig.~\ref{fig:fig2}(c).

\textit{Optimal decoupling}.--- For each bath oscillator Eq.~\eqref{eq:eom},
\begin{align}
    \dot{f}_k = i \omega_k' (f_k-f_{k0}u(t)),
\end{align} describes a particle in a moving harmonic potential, with position controlled by $u(t)$. Minimizing the residual population, furthermore, corresponds to minimizing the number of excitations, $\sum |f_k(t_f)|^2$, at the end of the movement. This same problem arises in different contexts, from industrial crane operation \cite{gonzalez-resines2017} to ion shuttling for quantum applications \cite{ding2020,martikyan2020, martikyan2020a,qi2021}. Different approaches have been introduced to optimize these problems, particularly in the context of shortcuts to adiabaticity \cite{ding2020,guery-odelin2019,bernardo2020,guery-odelin2023}. For a single harmonic oscillator, or a small finite set of frequencies \cite{martikyan2020a}, counteradiabatic driving or inverse engineering can be used to minimize excitations. However, these methods cannot be used for a continuum of oscillators. Robustness to frequency perturbations can be added \cite{guery-odelin2014,zhang2022,qi2021,espinos2022}, which suppresses excitations in a small range around a finite set of frequencies, but we find that oscillators outside these regions end up in highly excited states, while trying to cover a bigger frequency region ends up requiring a fine-tuned drive with very large amplitudes, which is unrealistic.

\begin{figure}[t]
  \includegraphics[width=\columnwidth]{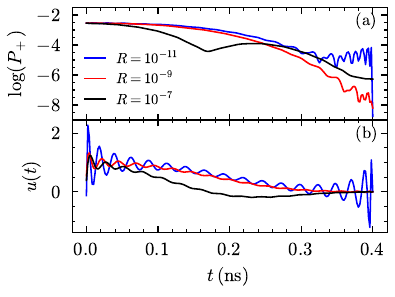}
\centering
\caption{ (a) Residual population of the qubit during decoupling, obtained for the optimal control given by the solution to the linear-quadratic regulator, given in (b). Computed for different values of the soft bound on the control $R$. }
\label{fig:lqr}
\end{figure}

To determine optimal decoupling protocols without assuming particular forms of coupling function $u(t)$ we therefore employ numerical optimal control, where the form of the equations of motion is also advantageous. Because the equations of motion in Eq.\eqref{eq:eom} are linear in both the displacement variables and the control, the system falls into the class of control-affine systems considered in classical optimal control \cite{kwakernaak1972}. 
If we consider a cost function with a soft bound on the drive
\begin{align}
    J = \sum_k f_k^2 + R \int_0^{t_f} dt\, u(t)^2,
\end{align}
we can identify the problem to solve as the linear-quadratic regulator, the most fundamental problem in classical optimal control. This has a known solution for the optimal control  \cite{kwakernaak1972}, given by
\begin{align}
    u(t) = - \sum_k F_{k}(t) f_{k}(t),
\end{align}
where $F_k(t)$ are functions that can be computed by backpropagation (see end matter). 

Fig.~\ref{fig:lqr} shows the residual populations, and the optimal controls, for different values of the soft bound $R=10^{-7},10^{-9},10^{-11}$. Imposing a moderate soft bound, $R=10^{-7}$, the optimal control displays a smooth decoupling with weak oscillations and a crossing of $u=0$. This results in a population of $P_+ \approx 10^{-6}$, similar to the results obtained with the smooth transition function (Fig.~\ref{fig:fig2}). Reducing the soft bound results in an improvement of the final qubit population of multiple orders of magnitude, but comes at the expense of a much more finely tuned control which is less practical. 

Furthermore, it is important to note that in practice the effect of the polaron discussed in this Letter is not the only effect limiting reset fidelity. At finite temperatures the fidelity will be limited by the thermal population of the excited state, which is $P_+ \approx 10^{-5}$ for a typical superconducting qubit~\cite{liu2015,schlor2019,hollister2022}, with $\omega_q/2\pi=5\,\mathrm{GHz}$ and $T=20\,\mathrm{mK}$. Thus, reducing the effect of the polaron beyond the level achievable using the smooth switching protocol is not relevant unless it is accompanied by the reduction of thermal and other sources of noise. 

\textit{Conclusion} --- Although system-environment correlations can limit the performance of quantum processes these limitations can be overcome through time-dependent control of the coupling or system.  For dissipative transmon reset, numerically exact simulations using a modified TEMPO algorithm showed that predicted limitations on speed and fidelity\ \cite{tuorila2019} are due to the formation of a polaron, which can be reversed by a smooth switch-off of the coupling. A treatment based on the time-dependent variational principle allowed us to show that smooth switching protocols achieve residual populations as low as $P_+\sim 10^{-6.5}$ at zero temperature. This reaches the levels required for quantum error correction, with a reset time of $\sim 10$ ns. While state-of-the-art experiments, which achieve $P_+\sim  10^{-3}$ in hundreds of nanoseconds~\cite{reed2010,zhou2021,yuan2023}, may be affected in practice by other issues such as $1/f$ noise, our analysis shows that transmon reset can be achieved with much higher fidelity, and much more rapidly, than expected, once these other effects are brought under control. Our methods could be applied more generally to show how the dynamics of system-environment correlations can be controlled in open quantum systems beyond the Born-Markov approximation.

\begin{acknowledgments}
We acknowledge funding from Taighde \'Eireann -- Research Ireland (21-FF-P/10142). 
\end{acknowledgments}

In order to meet institutional and research funder open access requirements, any accepted manuscript arising shall be open access under a Creative Commons Attribution (CC BY) reuse license with zero embargo.

Code and data supporting this article are openly available~\cite{data}.

\bibliography{landreset.bib,addmaterials.bib}

\clearpage

\appendix
\section*{End matter}

\textit{Time-dependent bath coupling in TEMPO}.--- 
Using the influence functional \cite{strathearn2018}, the evolution of the reduced density matrix is obtained as
\begin{align}
    \rho_{\mu_n} (t_n) =& \sum_{\{\mu\}}  \left( \prod_{k_1=1}^{n-1} [U_{k_1}]_{\mu_{k_1+1}\mu_{k_1}}\right) \nonumber \\
    &\cross \prod_{k_1=1}^n\prod_{k_2=1}^{k_1} I_{\mu_{k_1}\mu_{k_2}} \rho_{\mu_0}(t_0),
\end{align}
where $U_k$ are the system propagators and the influence functional blocks are given by
\begin{align}
    [I_{k_1 k_2}]_{\mu\nu} = \exp[-O^-_j(O^-_{j'} \Re[\eta_{k_1 k_2}]+iO^+_j \Im[\eta_{k_1 k_2}])],
\end{align}
where $O_j^\pm = [\hat{O},\, \cdot \, ]_{\pm,jj}$, and $\eta_{k k'}$ are integrals of the correlation function
\begin{align}
    \eta_{k_1 k_2} = \int_{t_{k_1}-1}^{t_{k_1}}d\tau_1\int_{t_{k_2}-1}^{t_{k_2}}d\tau_2 \, C(\tau_1,\tau_2).
\end{align}

The calculation above involves the compression of a tensor network of $n^2/2$ influence functional blocks, where $n$ is the total number os steps. For a time-independent bath coupling, the influence functional blocks only depend on the difference, $I_{\mu_{k_1} - \mu_{k_2}}$, such that only $n$ blocks have to be computed. When we introduce a global control on the coupling, $\tilde{g}_k(t) = g_k u(t)$, this is no longer true, and one needs to compute all $n^2/2$ blocks, which greatly increases computational time. This, however, can be avoided by realizing that the correlation function is proportional to the correlation function for a time-independent coupling, $u=1$,
\begin{align}
    C(\tau_1, \tau_2) = u(\tau_1)u(\tau_2)C^{u=1}(\tau_1- \tau_2).
\end{align}
This allows us to obtain the influence functional blocks as
\begin{align}
    [I_{k_1 k_2}]_{jj'} = ([I_{k_1-k_2}^{u=1}]_{jj'})^{u(t_{k_1})u(t_{k_2})},
\end{align}
where we assume a constant $u(t)$ within each time-step. Note that this is an element-wise exponentiation, and not a matrix exponentiation, which is a much simpler calculation. This allows us to introduce a time-dependent bath coupling to the TEMPO algorithm without significantly affecting its efficiency.

However, other algorithms have been developed recently which exploit time-translational symmetry \cite{link2024} to compress the influence functional much more efficiently than the TEMPO algorithm. These algorithms, by definition, cannot be extended to the time-dependent case as it breaks time-translational symmetry.

\textit{Time-dependent variational principle with the polaron ansatz}.--- 
In the time-dependent variational principle a manifold of ansatz states is defined, $\ket{\psi(\vec{x})}$, characterized by the real variational parameters $\vec{x}$, and the Schr\"odinger equation is projected onto the tangent space, such that evolution remains within the manifold \cite{hackl2020}. The projection leads to a set of equations of motion for the variational parameters,
\begin{align}
    \dot{x}_\mu B_{\mu\nu} = \partial_\nu  E.
\end{align}
where $E= \bra{\psi(\vec{x})}H\ket{\psi(\vec{x})}$, $B_{xy} = \partial_{y_k}A_{x_k} - \partial_{x_k}A_{y_k}$ is the geometric curvature and $A_{x} = i \bra{\psi} \partial_{x} \ket{\psi} $, the geometric connection.

For the spin-boson model, with
\begin{align}
    H = \frac{1}{2} \boldsymbol{h} \cdot \boldsymbol{\sigma}+ \sum_k \omega_k b_k^\dagger b_k + \frac{\sigma_z}{2} \sum_k g_k (b_k^\dagger + b_k),
\end{align}
we use the polaron ansatz,
\begin{align}
    \ket{\Psi(\boldsymbol{{f}})} = \frac{1}{\sqrt{2}} \left(\ket{\uparrow,\boldsymbol{f}} - \ket{\downarrow,-\boldsymbol{f}} \right),
\end{align}
where $\ket{ \boldsymbol{f}} =  \Pi_k D_k( f_k)\ket{0}$, and $D_k(\pm f_k)$ are the displacement operator for each oscillator, labeled by $k$. They are unitary, $D(\alpha)D(\beta) = D(\alpha+\beta)$, act on the bosonic operators as a shift, $ D^\dagger(\alpha) b D(\alpha) = b + \alpha$, and the resulting coherent states are not orthonormal, $\bra{\beta}\ket{\alpha} = e^{(2\beta ^* \alpha - \abs{\beta}^2-\abs{\alpha}^2)/2}$. We can then use the bath displacements as variational parameters to approximate the dynamics of the system near the polaron state. Since $f_k$ is complex and the parametrization is non-holomorphic, we need to consider its real and imaginary parts, $f_k = x_k + i y_k$, as independent variational parameters.

To calculate the expectation value of the Hamiltonian we compute first the overlaps
\begin{align}
    & \bra{\uparrow,f}H \ket{\uparrow,f} =  \frac{h_z}{2} + \sum_k \omega_k \abs{f_k}^2 + \sum_k g_k(f_k+f_k^\ast)/2 \nonumber \\
    &\bra{\uparrow,f}H \ket{\downarrow,-f} =   \frac{h_x- i h_y}{2}e^{-2\sum_k \abs{f_k}^2} \nonumber \\
    &\bra{\downarrow,-f}H \ket{\uparrow,f} =  \frac{h_x+i h_y}{2} e^{-2\sum_k \abs{f_k}^2}  \nonumber \\
    & \bra{\downarrow,-f}H \ket{\downarrow,-f} =   \bra{\uparrow,f}H \ket{\uparrow,f} - h_z ,
\end{align}
resulting in
\begin{align}
    E =& -\frac{1}{2}h_x e^{-2\sum_k \abs{f_k}^2} + \sum_k \omega_k \abs{f_k}^2 +\frac{1}{2}\sum_k g_k  ( f_k +f_k^*),
\end{align}
with derivatives
\begin{align}
    &\partial_{x_k}E(\boldsymbol{f}) = 2 x_k h_x e^{-2\sum_k (x_k^2+y_k^2)} +2 \omega_k x_k + g_k \nonumber \\
    &\partial_{y_k}E(\boldsymbol{f}) = 2 y_k h_x e^{-2\sum_k (x_k^2+y_k^2)} +2 \omega_k y_k .
\end{align}
 We continue by computing the derivatives of the variational state,
\begin{align}
    &\partial_{x_k}\ket{\psi(\boldsymbol{{f}})} = \frac{1}{\sqrt{2}}(b_k^\dagger - b_k)\left(\ket{\uparrow,\boldsymbol{f}}+\ket{\downarrow,-\boldsymbol{f}}\right) \nonumber \\
    &\partial_{y_k}\ket{\psi(\boldsymbol{{f}})} = i\frac{1}{\sqrt{2}}(b_k^\dagger + b_k)\left(\ket{\uparrow,\boldsymbol{f}}+\ket{\downarrow,-\boldsymbol{f}}\right),
\end{align}
and the Berry connections,
\begin{align}
    A_{x_k} =& i\frac{1}{2} \bra{\uparrow,\boldsymbol{f}} (b_k^\dagger - b_k)\ket{\uparrow,\boldsymbol{f}} \nonumber \\
    &-i\frac{1}{2}\bra{\downarrow,-\boldsymbol{f}} (b_k^\dagger - b_k)\ket{\downarrow,-\boldsymbol{f}}  \nonumber \\
    =& 2 y_k,
\end{align}
and $A_{y_k} = -2x_k$, resulting in the Berry curvature $ B_{x_k y_k} = 2$. Altogether we obtain the equations of motion
\begin{align}
    \dot{f}_k = if_k(h_x e^{-2\sum_{k}\abs{f_k}^2}+\omega_k) + \frac{i}{2} g_k.
\end{align}
It is interesting to note that, in the weak coupling limit, the exponential vanishes and the equations of motion become independent of each other, greatly simplifying their solution.

\textit{Linear-quadratic regulator}.--- 
Consider a discrete-time linear control-affine system \cite{kwakernaak1972}, given by the equations of motion
\begin{align}\label{eq:lqr_eom}
    x_{p,t+1} = \sum_{q=1}^N A_{pq} x_{q,t} + \sum_{n=1}^M B_{pn}u_{n,t},
\end{align}
where $\vec{x} \in \mathbb{R}^N$ and $\vec{u} \in \mathbb{R}^M$. Consider as well a quadratic cost function
\begin{align}
    J =& \sum_{pq} x_{p,t_f} Q_{t_f,pq} x_{q,t_f} + \sum_t x_{p,t} Q_{pq}x_{q,t} \nonumber \\
    &+ \sum_{nm}u_{n,t}R_{nm}u_{m,t} + 2 \sum_{pn} x_{p,t} N_{pn} u_{n,t}.
\end{align}
The control protocol that minimizes the cost function is known to be
\begin{align}
    u_{n,t} = -\sum_{p} F_{np,t} x_{p,t},
\end{align}
where
\begin{align}
    F_t = (R+B^T P_{t+1}B)^{-1}(B^T P_{t+1} A +N^T),
\end{align}
and $P_t$ is obtained by backpropagating the Ricatti equation
\begin{align}
    P_{t-1} =& -(A^T P_t B + N)(R+B^TP_t B)^{-1}(B^T P_t A + N^T) \nonumber \\
    &+A^T P_t A +Q
\end{align}
from the initial value $P_{t_f} = Q_{t_f}$.

Consider again the EOM for the bath displacements in the spin-boson model, in Eq.\eqref{eq:eom}, with a time-dependent coupling
\begin{align}
    \dot{f}_k = i f_k (\omega_q  + \omega_k)  + i\frac{1}{2}u(t)g_k.
\end{align}
We discretize time by integrating the EOMs over a time-step, considering the controls to be constant during the time-step,
\begin{align}
    f_{k,t+1} = e^{i \omega_k^q \delta_t}f_{k,t}+u_t g_k \frac{e^{i\omega_k^q \delta_t}-1}{2\omega_k^q}.
\end{align} 
In order to express it in the form of Eq.\eqref{eq:lqr_eom} we decompose the displacements into real and imaginary parts, $f_k = f_k' + i f_k''$, with
\begin{align}
    &f'_{k,t+1} = \cos(\omega_k^q \delta_t) f'_{k,t} - \sin(\omega_k^q \delta_t) f''_{k,t} + u_t g_k \frac{\cos(\omega_k^q \delta_t)-1}{2\omega_k^q}, \nonumber \\
    &f''_{k,t+1} = \sin(\omega_k^q \delta_t) f'_{k,t} + \cos(\omega_k^q \delta_t) f''_{k,t} + u_t g_k \frac{\sin(\omega_k^q \delta_t)}{2\omega_k^q}, 
\end{align}
Considering the new variable vector $\vec{x} = (f'_1,f''_1,f'_2,f''_2,...)$, we can rewrite the EOMs as
\begin{align}\label{eq:lqr}
    &x_{k\mu,t+1} = A_{kp,\mu\nu} x_{p\nu,t} + B_{k\mu}u_t,
\end{align} 
where
\begin{align}
    &A_{kp,\mu\nu} = \delta_{kp}\mqty(\cos(\omega_k^q \delta_t) & -\sin(\omega_k^q \delta_t) \\ \sin(\omega_k^q \delta_t) & \cos(\omega_k^q \delta_t))_{\mu\nu}, \\
    &B_{k\mu} = \frac{g_k}{2\omega_k^q} \mqty(\cos(\omega_k^q \delta_t)-1 \\ \sin(\omega_k^q \delta_t) )_\mu .
\end{align}
We also consider the cost function
\begin{align}
    J = \sum_{k\mu} x_{k\mu,t_f}^2  + R \sum_t u_t ^2,
\end{align}
where the first term minimizes the displacement in the final state, i.e. its infidelity with respect to the ground state, and the second term is a soft bound that ensures the control remains finite. We therefore have $Q_{t_f} = I, Q=0,N=0$, and the Ricatti equation can be solved to obtain an optimal control.

\end{document}